\documentstyle[11pt,newpasp,twoside,epsf]{article}
\def\edcomment#1{\iffalse\marginpar{\raggedright\sl#1\/}\else\relax\fi}
\reversemarginpar

\begin{document}
\title{The bulge globular cluster NGC 6553: Observations with HST's WFPC2, STIS and NICMOS}
 \author{Sylvie F. Beaulieu, Gerard F. Gilmore, Rachel A. Johnson \& Stephen J. Smartt}
\affil{Institute of Astronomy, Cambridge, UK, beaulieu@ast.cam.ac.uk} 
\author{Nial Tanvir}
\affil{Department of Physical Science, University of Hertfordshire, UK}
\author{Basilio Santiago}
\affil{Universidade Federal do Rio Grande do Sul, Porto Alegre, Brazil}

\begin{abstract}
As part of a large HST project to study the formation and
evolution of rich star clusters in the Large Magellanic Cloud,
we present results for the Galactic bulge metal-rich globular
cluster NGC6553. HST observations using WFPC2, NICMOS and STIS 
were obtained for this cluster. The primary reason for studying 
NGC6553 is to transform our NICMOS
and STIS magnitudes directly into absolute magnitudes. This is
particularly important for determining the low mass end of 
the IMF. NGC 6553 was chosen because its metallicity,
[Fe/H]$~ \approx -0.2$, is representative of the metallicities
of the young and intermediate age LMC clusters.
\end{abstract}

\keywords{Galaxy: bulge globular clusters: individual: NGC 6553 - photometry: HST}

\section{Introduction}
The Galactic bulge globular cluster NGC 6553 was observed as part of
a large HST project to study the formation and evolution of rich star clusters
in the Large Magellanic Cloud. A total of 95 orbits were awarded to this
project in Cycle 7, using WFPC2, STIS (in imaging mode), and NICMOS to
obtain multi-wavelength photometry in eight clusters with ages $\sim 10^7,
10^8, 10^9$ and $10^{10}$ years. Data acquisition was completed in November 1998.
Early results have been presented by Beaulieu et al. (1998), Elson et al. (1998c), 
and Johnson et al. (1998,1999).
The results of a pilot study of one of the clusters in the sample
are described in Elson et al. (1998a,b).

NGC 6553 was observed for calibration purposes, in order to transform
NICMOS and STIS magnitudes directly into absolute magnitudes on a
standard system.
Beyond calibration, our data allow us to investigate various properties
of the cluster. Clusters like NGC 6553 are of particular interest in
that they serve as tracers of the formation and evolution of the bulge
component of the Milky Way. In this regard, any spreads in age or
metallicity among bulge clusters provide clues concerning the enrichment
history of the bulge, the timescale for its formation, and the time of
its formation relative to the halo and disk. Their dynamically vulnerable
location also allows the effect of tidal shocking on the structure and
stellar content of the clusters to be explored. Our WFPC2 data allow us
to probe the stellar content of NGC 6553 well below its main-sequence
turnoff, and determine accurate values of the cluster's distance and
reddening. Our STIS data allow us to determine a deep luminosity function (LF)
which may be compared with LFs in other Galactic globular clusters.

We are investigating differential reddening as the explanation for all the
apparent peculiarities of this cluster. The peculiarities include the 
red giant branch (RGB) bump, the tilted horizontal branch (HB),
the apparent ''second turnoff" and the unusual LF 
consistent with these studies.

\section{WFPC2 CMDs}

The main features of the colour-magnitude diagrams (CMDs) include the blue 
population brightwards of the main-sequence turnoff, the tilted HB and the clump 
below the HB. A 12 Gyr isochrone for [Fe/H]$=-0.4$, reddening
$\rm E(B-V)=0.7$, and distance modulus $\rm (m-M)_0=13.7$ fit best the data (Fig 1).
Also visible in the CMDs of WFPC2 is an apparent ''second turnoff", an excess
of stars near $\rm (V,V-I) = (21,2)$. 
The second fainter turnoff is particularly
prominent in chips WF4 and fainter in the PC and WF2 (where WF3 is the cluster 
core). It is not understood yet what this turnoff represents, it could be
associated with the Galactic bulge, or could be the result of patchy
reddening (see next subsection). 

\subsection{A tilted horizontal branch}

It has been noted by previous authors that the HB in
NGC 6553 is far from horizontal. Its range in V spans $\sim 0.5$ mag.
Some authors have attributed this tilt to differential reddening across
the face of the cluster (cf. Guarnieri et al. 1998). Others suggest that
metal line blanketing can produce a tilting of the HB 
(cf. Ortolani et al. 1990). 

\subsection{RGB bump}

The HB shows up prominently, peaking at $\rm V_{555}=16.6$
(Fig 1). A second peak is clearly visible $\sim 0.9$ magnitudes below the
HB. This has been discussed, for example, by Sagar et al.
(1999) and attributed to a phase of stellar evolution where the star 
becomes fainter. Other authors have suggested that it could also be
due to the superposition on the cluster RGB of background HB stars,
however this is not consistent with the redward shift of the background RGB.

\section{STIS-LP LF}

Fig 2 shows the STIS-LP LF for NGC 6553, both raw and
corrected for incompleteness. In the bottom panel, the instrumental
magnitude has been transformed to an absolute magnitude in the STIS-LP
passband, adopting an aperture correction of $-0.5$ mag, a zero point
of 23.4 mag, a reddening of $\rm E(B-V)=0.7$ and an absolute distance modulus
of $\rm (m-M)_0=13.7$. The reddening and distance information are from
Guarnieri et al. (1998). The value of $\rm E(B-V)$ has been transformed to 
an absorption in the STIS-LP passband by $\rm A_{R(STIS)} = 2.505 \cdot E(B-V)$.
This LF differs from that of most globular clusters (cf. Elson et al. 1998c),
possibly because of significant uncorrected extinction.

\section{NICMOS2 CMD}

Fig 3 shows the CMD of $\rm V_{555}$ vs ($\rm V_{555}-H_{160}$) for the short 
exposure. A 12 Gyr isochrone for [Fe/H]$=-0.4$, reddening $\rm E(B-V) = 0.7$
and distance modulus $\rm (m-M)_0 = 13.7$ is overlayed but clearly does not
represent the data. The considerable width of the main sequence, though
less than in Fig 1, suggests important patchy extinction which may also
be responsable for the difficulty in fitting isochrones.  



\begin{figure}
\plotone{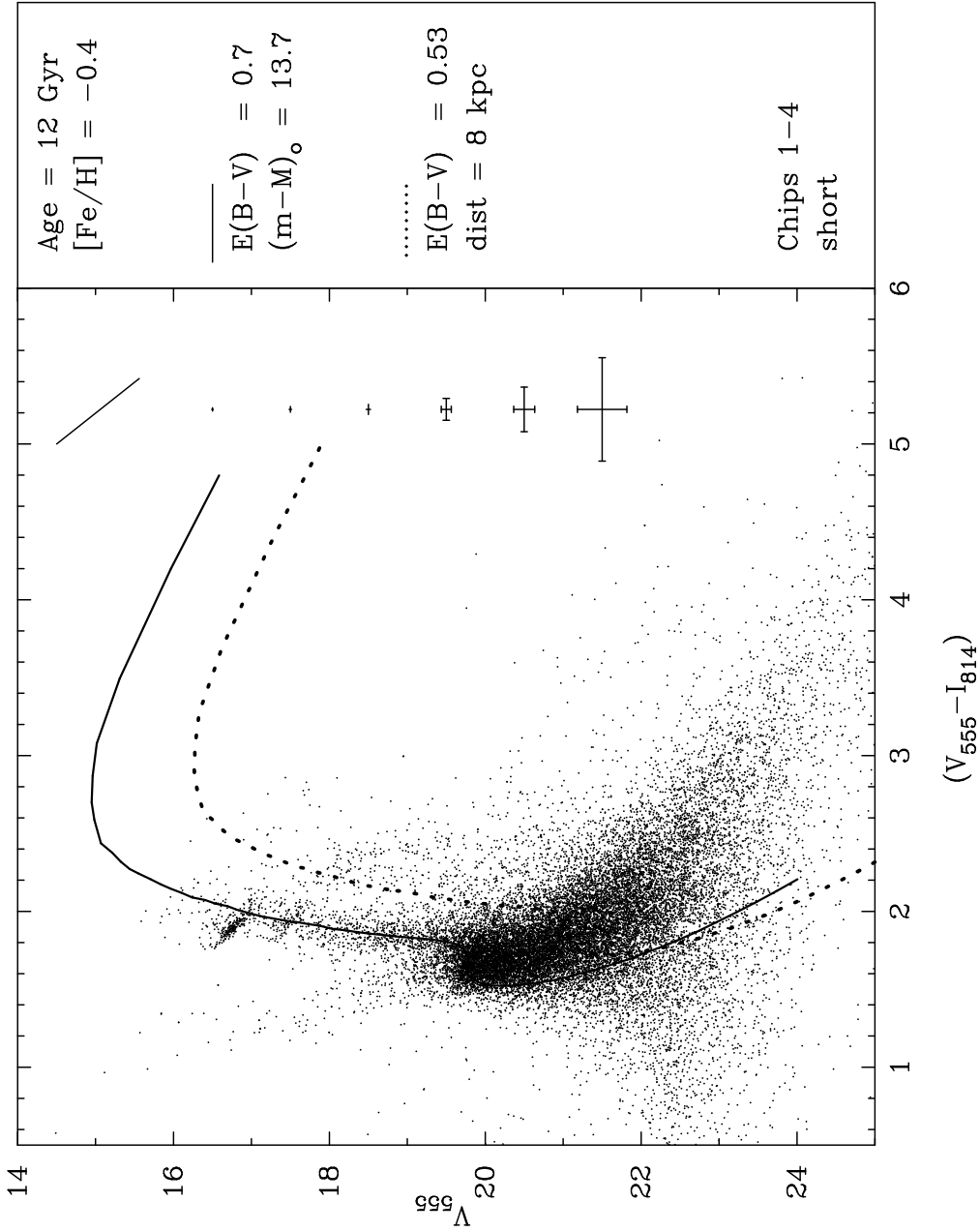}
\caption{CMD for the short exposure observations of NGC 6553.
A 12 Gyr isochrone (Bertelli et al., 1994; Wortley, 1998) for [Fe/H]$=-0.4$, 
reddening $\rm E(B-V)=0.7$,
and distance modulus $\rm (m-M)_0=13.7$ is superposed. The dashed isochrone
represents the background bulge stars and has [Fe/H]$=-0.4$, $\rm E(B-V)=0.53$,
and distance 8 kpc.. The diagonal line indicates the reddening vector.
The saturation limit is at $\rm V_{555} \approx 15.5$. The very
considerable width of the main sequence and the tilted HB suggest
this CMD is strongly affected by patchy extinction.}
\end{figure}

\begin{figure}
\plotone{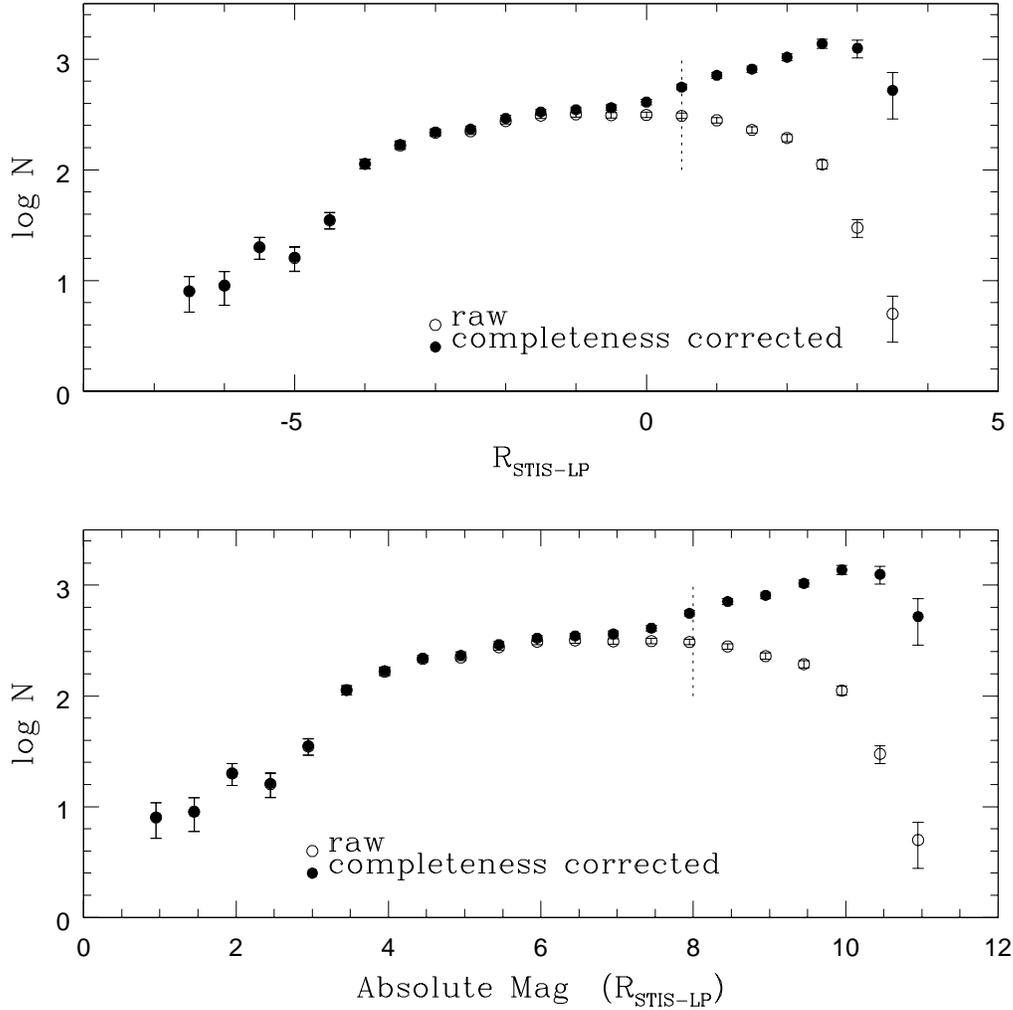}
\caption{Instrumental and absolute luminosity functions derived
from STIS for a field $\sim 1$ arcmin from the centre of NGC 6553.
In the bottom panel, absolute magnitude $\rm = R_{STIS} -0.5 + 23.4 - 13.7 - 1.75$.
Poisson error bars are shown. Open circles are from raw counts and
filled circles have been corrected for incompleteness. The vertical
dashed line indicates the 50\% completenss limit. The LF may still be substantially
affected by differential extinction.}
\end{figure}

\begin{figure}
\plotone{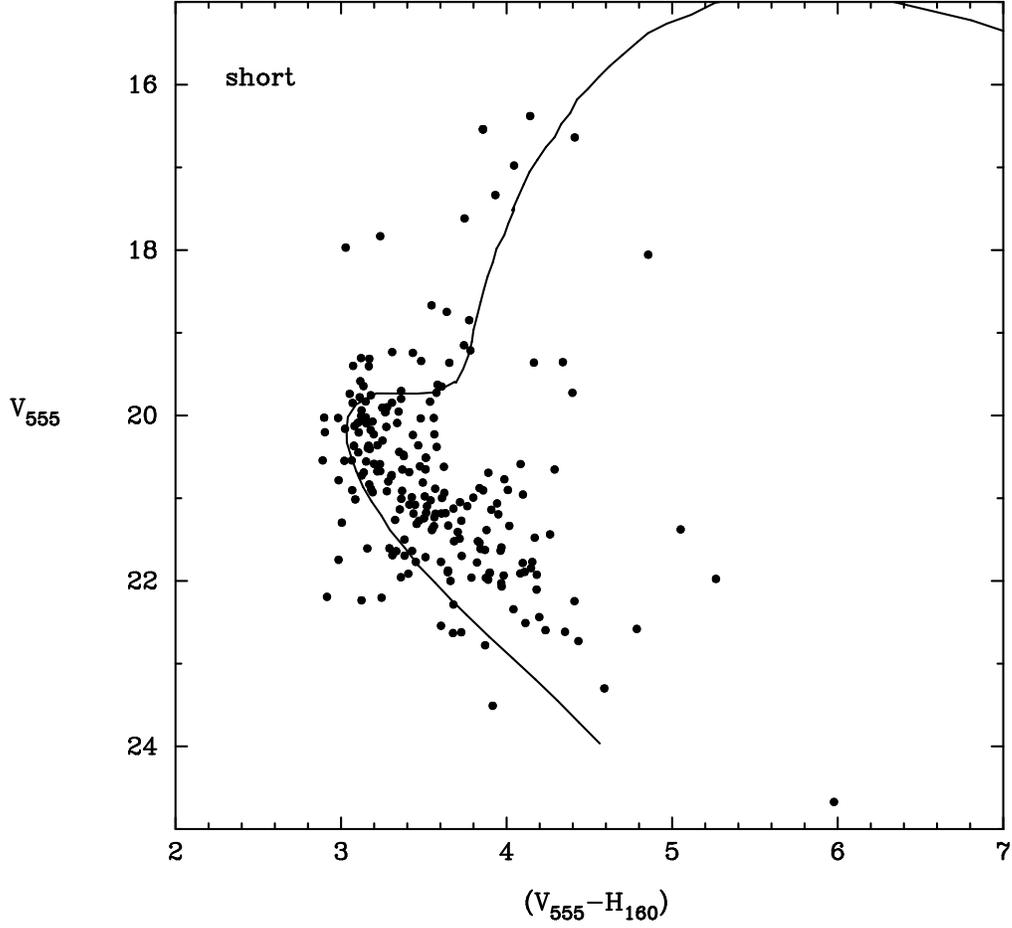}
\caption{CMD of $\rm V_{555}$ vs ($\rm V_{555}-H_{160}$) for the short exposure
observations of NGC 6553. A 12 Gyr isochrone for [Fe/H]$=-0.4$, reddening
$\rm E(B-V)=0.7$ and distance modulus  $\rm (m-M)_0=13.7$ is superposed.}
\end{figure}

\end{document}